\documentclass[conference]{IEEEtran}
\IEEEoverridecommandlockouts
\usepackage{cite}
\usepackage{amsmath,amssymb,amsfonts}
\usepackage{graphicx}
\usepackage{textcomp}
\usepackage{xcolor}
\usepackage[english]{babel}
\usepackage{blindtext}
\usepackage{algorithm}
\usepackage[noend]{algpseudocode}
\usepackage{algpseudocode}
\usepackage{graphicx, caption, subcaption}
\usepackage{multirow}
\usepackage[T1]{polski}
\usepackage[normalem]{ulem}
\usepackage{tikz}
\def\BibTeX{{\rm B\kern-.05em{\sc i\kern-.025em b}\kern-.08em
    T\kern-.1667em\lower.7ex\hbox{E}\kern-.125emX}}

\IEEEoverridecommandlockouts\IEEEpubid{\makebox[\columnwidth]{979-8-3503-5171-2/24/\$31.00 $\copyright$2024 IEEE \hfill}\hspace{\columnsep}\makebox[\columnwidth]{ }}

\newcommand\copyrighttext{%
  \footnotesize \textcopyright 2024 IEEE. Personal use of this material is permitted. Permission from IEEE must be obtained for all other uses, including reprinting/republishing this material for advertising or promotional purposes, collecting new collected works for resale or redistribution to servers or lists, or reuse of any copyrighted component of this work in other works.}

\begin{document}

\newcommand{\etal}[0]{\textit{et al.}}
\newcommand{\Input}{\hspace*{\algorithmicindent} \textbf{Input: }}
\newcommand{\Output}{\hspace*{\algorithmicindent} \textbf{Output: }}

\newcommand{\blob}{cluster\xspace}
\newcommand{\blobs}{clusters\xspace}
\newcommand{\Blob}{Cluster\xspace}
\newcommand{\Blobs}{Clusters\xspace}
\newcommand{\blobbed}{clustered\xspace}
\newcommand{\blobbing}{clustering\xspace}
\newcommand{\Blobbing}{Clustering\xspace}
\newcommand{\bRt}{QuARC\xspace}
\newcommand{\quarc}{QuARC\xspace}
\newcommand{\new}[1]{{\color{blue}{#1}}}
\newcommand{\del}[1]{{\color{purple}\sout{#1}}}

\newcommand*{\figuretitle}[1]{{\centering \textbf{#1}\par\medskip}}

\newcommand\copyrightnotice{%
    \begin{tikzpicture}[remember picture,overlay]
    \node[anchor=south,yshift=10pt] at (current page.south) {\parbox{\dimexpr0.75\textwidth-\fboxsep-\fboxrule\relax}{\copyrighttext}};
    \end{tikzpicture}%
    \vspace{-1em}
}

\title{Efficient Routing on Quantum Networks using Adaptive Clustering}

\author{\IEEEauthorblockN{Connor Clayton}
\IEEEauthorblockA{\textit{University of Maryland} \\
cbclayto@cs.umd.edu}
\and
\IEEEauthorblockN{Xiaodi Wu}
\IEEEauthorblockA{\textit{University of Maryland} \\
xwu@cs.umd.edu}
\and
\IEEEauthorblockN{Bobby Bhattacharjee}
\IEEEauthorblockA{\textit{University of Maryland} \\
bobby@cs.umd.edu}
}

\maketitle
\copyrightnotice

\begin{abstract}
    We introduce QuARC, Quantum Adaptive Routing using Clusters, a novel
clustering-based entanglement routing protocol that leverages
redundant, multi-path routing through multi-particle projective
quantum measurements to enable high-throughput, low-overhead,
starvation-free entanglement distribution.  At its core, QuARC
periodically reconfigures the underlying quantum network into clusters
of different sizes, where each cluster acts as a small network that
distributes entanglement across itself, and the end-to-end
entanglement is established by further distributing between clusters.
QuARC does not require a-priori knowledge of any physical parameters,
and is able to adapt the network configuration using static topology
information, and using local (within-cluster) measurements only.  We
present a comprehensive simulation-based evaluation that shows QuARC
is robust against changes to physical network parameters, and
maintains high throughput without starvation even as network sizes
scale and physical parameters degrade.

\end{abstract}

\section{Introduction}

Recent papers~\cite{pant-routing, Q-CAST,Towsley,multiplexed-Towsley} have
defined a framework for studying entanglement routing in quantum networks, leading to the
development of several promising schemes. 
To date, the predominant family of concurrent
quantum routing algorithms~\cite{Q-CAST, fragmentation-Q-CAST, EFiRAP,REPS,zeng2022multi} relies on entanglement
swapping for entanglement distribution, often assuming that
entanglement creation and swapping probabilities are known a-priori.
Newer designs~\cite{Towsley,multiplexed-Towsley,alg-n-fusion,multi-partite-routing} use multi-particle
projective quantum measurements, referred to as \textit{fusions}, that take
advantage of multiple concurrent paths to enable entanglement over
larger topologies.  Both of these approaches have certain benefits
and drawbacks.

Multi-path entanglement distribution protocols can provide very high
throughput~\cite{Q-CAST,REPS,alg-n-fusion,fragmentation-Q-CAST}, in
terms of number of simultaneous successful entanglements, under
favorable conditions.  These protocols, however, are increasingly
ineffective over larger network diameters, and 
rely heavily on sufficiently favorable physical conditions that must be known a-priori.
Entanglement swapping protocols are also likely to ``starve''
longer-distance/larger-hop entanglement requests, as their
effectiveness decreases rapidly with distance and hop count.  More
fundamentally, we show that the efficacy of these protocols is highly sensitive to the values of physical parameters and network size.

Fusion-based protocols, on the other
hand, were designed to generate distance-independent entanglement~\cite{Towsley,multiplexed-Towsley}, and
are largely agnostic to network diameter.  As long as the underlying
topology provides sufficient connectivity, and the physical parameters
are better than a critical threshold, these protocols can successfully
provide end-to-end entanglement.  These protocols, however,
often cannot provide high entanglement throughput, either because they
fail to take full advantage of multi-path routing~\cite{alg-n-fusion} or they are
restricted to servicing a single source-destination pair at a
time~\cite{Towsley,multiplexed-Towsley}.

We introduce \quarc, Quantum Adaptive Routing using Clusters, a new
quantum routing protocol that relies on adaptive \blobbing to address
these shortcomings of previous protocols.  \quarc does not require any
time-varying global knowledge, nor does it assume the knowledge of
physical parameters such as entanglement generation rates on any edge.
Instead, \quarc uses a local measurement-based process to reconfigure
the network periodically into fusion domains (or \blobs), over
which entanglement requests are served.  Depending on (changing)
physical parameters and entanglement request distribution, \quarc
creates larger or smaller \blobs, which are split or merged over time
to maintain high performance with respect to throughput and request
fulfillment latency.

\quarc is the first quantum routing protocol based on clustering, which is designed to strike a balance between providing
high overall entanglement throughput and sustaining high success-rates
for long distance entanglement requests.
We evaluate \quarc using simulations, and study the effectiveness and
robustness of \blobbing across a wide range of simulated physical
parameters and network sizes.  Importantly, we show that \quarc largely
maintains the distance-independent nature of previous fusion-based
protocols~\cite{Towsley,multiplexed-Towsley}, in that performance does not suffer (comparatively) as
physical parameters degrade or network diameters increase.
Simultaneously, we show that the performance of the best known multi-path entanglement routing protocols collapse (often to near-zero) under the same physical parameter shifts.

The rest of this paper is structured as follows: Section~\ref{bg}
provides a background on entanglement routing, including a discussion of
related work.  Section~\ref{design} presents \quarc's design, followed
by simulation evaluations in Section~\ref{eval}.  We discuss \quarc's
limitations and avenues for future work, as well as a realistic distributed implementation in
Section~\ref{discuss}; we conclude in Section~\ref{conc}.

\section{Background and Related Work\label{bg}} 

The use of traditional packet-switching in quantum networks is
largely precluded by two fundamental quantum mechanical facts: general quantum
messages cannot be copied~\cite{no-cloning} (the so called
``no-cloning theorem''), and, without error correction, the
information contained within these messages decays rapidly over
time. Additionally, quantum information decays exponentially with
distance~\cite{pirandola-repeaterless-limits}, meaning that directly sending a long-distance quantum
message fails with high probability, and attempting to resend that
message is (in general) impossible due to the no-cloning theorem.

Instead, future large-scale quantum networks are expected to operate
by \textit{entanglement distribution}. Entanglement distribution uses
two-particle entangled states known as \textit{Bell pairs}. When two
(distant) parties A and B are each in possession of one of the two
particles in a Bell pair, we say that A and B \textit{share
entanglement}. This shared entanglement is a resource which A can use
to send a single quantum bit, or \textit{qubit}, to B through a process known
as \textit{quantum teleportation}~\cite{teleportation}. The benefit of entanglement
distribution (over direct transmission) is that, unlike general
qubits, Bell pairs can be regenerated, the entanglement distribution attempt
repeated until successful.

The success rate of sending a Bell state particle through fiber also
decays exponentially with distance.  Long distance networks therefore will
rely on intermediate nodes known as 
\textit{quantum repeaters}~\cite{wehner-road-ahead, pirandola-repeaterless-limits}.  A long-distance entanglement between A and
B can be formed via an intermediate repeater C as follows:  A and B each separately 
distribute entanglement with C.  C then completes the long-distance
A-B entanglement using \textit{entanglement swapping}~\cite{swapping-1, swapping-2, swapping-3, swapping-4}, essentially
teleporting the two entanglements (A-C, C-B) locally.  This process
generalizes to multiple intermediaries, and such a sequence of
entanglement swaps can be performed in parallel, which reduces latency
and minimizes decoherence.  
Recent experiments have realized the foundational elements of multi-node entanglement swapping quantum networks~\cite{3-node-network-wehner, non-neighboring-teleportation},
and more limited quantum networks based on so-called ``trusted repeaters'' have been
deployed at metropolitan scales and beyond for over a decade~\cite{cambridge-qnet,
  chinese-trusted-repeater-qnet-1200km,
  chinese-trusted-repeater-qnet-4600km, tokyo-qkd-network,
  vienna-qkd-network, hefei-trusted-repeater-qnet-46-nodes}.

\paragraph{Network Model} This setting forms the basis for studying quantum networks: the
network is modeled as a graph, where each node contains some number of
\textit{memory qubits}, and each edge contains one or more \textit{quantum
channels}. The number of quantum channels in an edge is known as that
edge’s \textit{width}. When two adjacent nodes each assign a qubit to a
channel $c$, that channel can attempt to generate entanglement between
those two qubits; this entanglement generation succeeds with
probability $p_c$. We call a successful entanglement over a quantum
channel a \textit{link}.  Previous works assume that $p_c$ is entirely
dependent on the length $L$ of $c$ by a network constant $\alpha$, i.e.,
$p_c=e^{-\alpha L}$; however, we note that $p_c$ may be affected by additional
factors (e.g., cable splices~\cite{boston-qnet}) and may also be time-varying due to,
e.g., temperature change, wind speed, quantum device drift and
re-calibration~\cite{boston-qnet, NIST-qnet-metric, Fang_2023}.

\paragraph{Routing Protocols using Entanglement Swapping}
This basic network model has been used to study physical limits of
entanglement generation, often focusing on specific physical
topologies.  
Pirandola \etal{} uncover the fundamental limits of
repeaterless quantum
communication~\cite{pirandola-repeaterless-limits} as well as upper
bounds on the transmission rates of arbitrary repeater-assisted
quantum communication schemes \cite{pirandola-end-to-end-capacity}.
Van Meter \etal{} \cite{dijkstra-routing} adapt classical path
selection methods to account for quantum resource utilization.
\cite{optimal-RED, wehner-LP-routing} formulate entanglement
distribution as linear programming problems. \cite{decentrailized-hierarchical-routing} propose a decentralized, hierarchical routing scheme.  Numerous works analyze
routing on special topologies such as chains, rings, grids, and trees
\cite{schoute-routing, caleffi-routing, das-routing, pant-routing,
  wehner-distributed-routing, li2021effective, fat-tree}.

Expanding on these fundamental ideas, entanglement routing algorithms
have been developed for the more general setting in which concurrent
source-destination pairs must be serviced on an arbitrary network
topology.  Shi and Qian~\cite{Q-CAST} developed the first of these,
proposing a comprehensive model of the entanglement routing problem
and corresponding algorithms for throughput maximization.  Zhang
\etal{}~\cite{fragmentation-Q-CAST} extend~\cite{Q-CAST} to more efficiently utilize local link state
information, and Zhao and Qiao~\cite{REPS} demonstrate that global
access to link state information can also improve throughput over
\cite{Q-CAST}.  \cite{request-scheduling, zeng2022multi} consider
request scheduling for improved fairness.  Several works examine
entanglement routing in various related settings, including under
noisy link generation~\cite{EFiRAP}, time
multiplexing~\cite{subexponential, opportunistic-routing}, and
non-homogeneous time slots~\cite{non-uniform-time-slots}.

\begin{figure*}[t]
\centering
    \begin{subfigure}[t]{.315\textwidth}
        \figuretitle{Single-path Routing}
        \includegraphics[width=\textwidth]{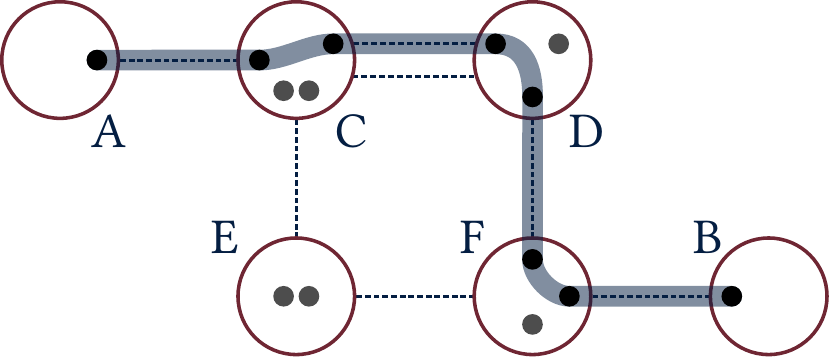}
        \caption{Single path selection.}
        
        \par\bigskip
        
        \includegraphics[width=\textwidth]{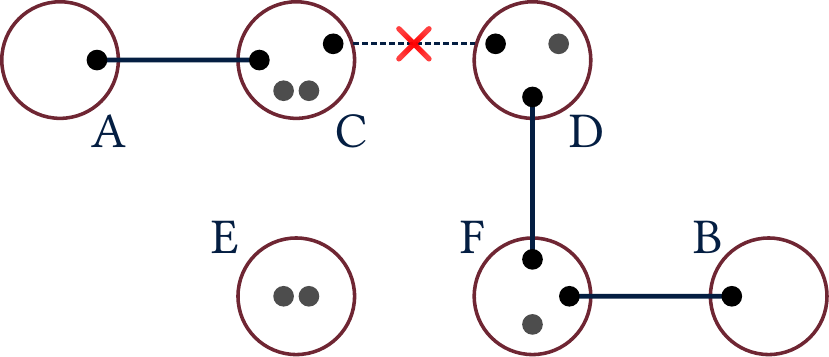}
        \caption{S-D entanglement fails due to single link failure}
    \end{subfigure}
    \hspace*{.01\textwidth}
    \begin{subfigure}[t]{.315\textwidth}
        \figuretitle{Multi-path Entanglement Swapping}
        \includegraphics[width=\textwidth]{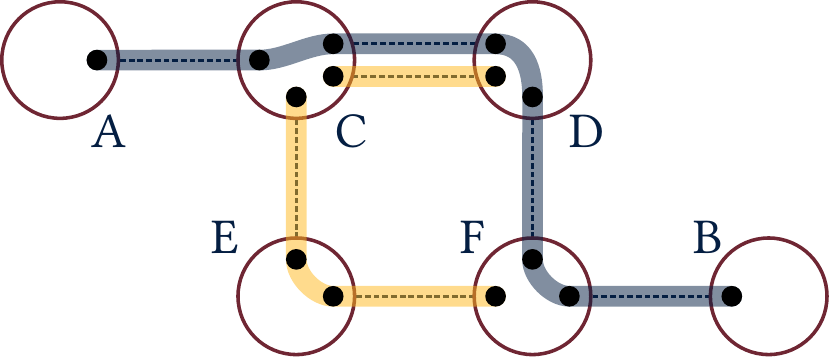}
        \caption{Path selection with recovery paths}

        \par\bigskip
        
        \includegraphics[width=\textwidth]{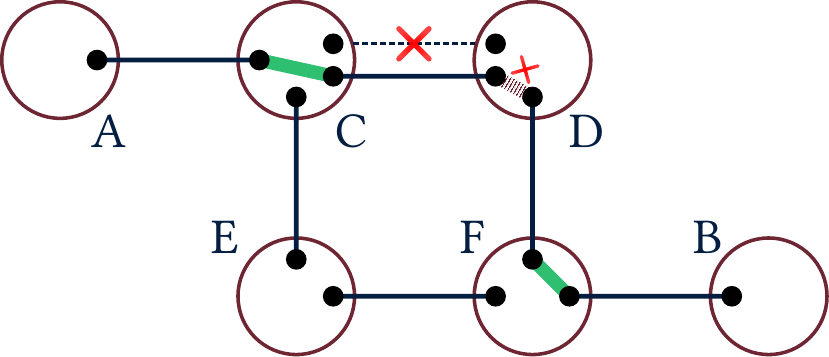}
        \caption{Recovery path averts failed link, but a failed swap
          prevents entanglement distribution.}
    \end{subfigure}
    \hspace*{.01\textwidth}
    \begin{subfigure}[t]{.315\textwidth}
        \figuretitle{Multi-path Routing with Fusions}
        \includegraphics[width=\textwidth]{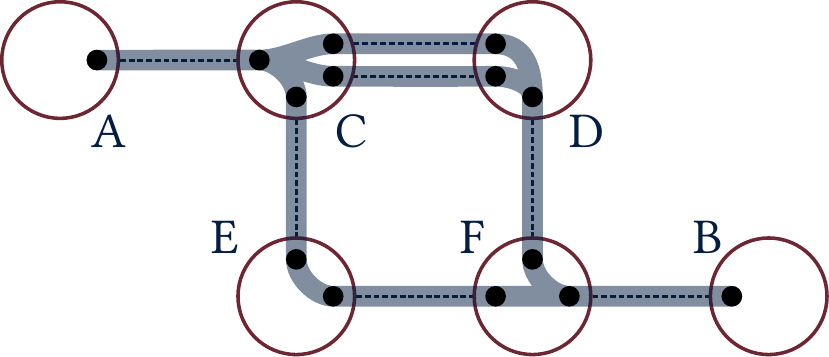}
        \caption{Fusion: all possible paths are selected.}

        \par\bigskip

        \includegraphics[width=\textwidth]{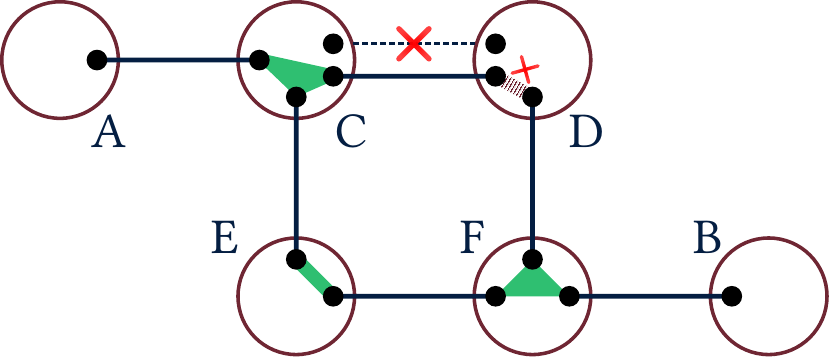}
        \caption{Each node performs a fusion on successfully entangled qubits. Entanglement distribution succeeds along the lower path.}
    \end{subfigure}

    \bigskip

    \includegraphics[width=\linewidth]{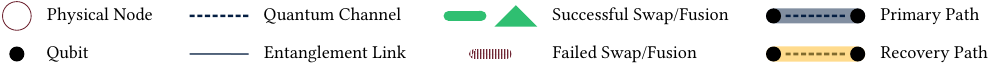}

\caption{Comparison of different quantum routing
  approaches.}
\label{fig:multi-path-routing}
\end{figure*}

\subsection{Fusion-based Entanglement Routing}

Recently, a generalization of entanglement swapping which acts on more
than two qubits has been studied in the context of quantum network routing~\cite{pant-routing,Towsley, multiplexed-Towsley, n-fusion-noise, alg-n-fusion, multi-partite-routing}. Known
as \textit{n-fusion}, or simply fusion, this operation acts on $n$
qubits and ``fuses'' the entanglement stored in these qubits’ links
into a larger multi-party entangled state (known as a GHZ state).
To illustrate this effect, consider a node $v$ which shares entanglement with $n$ neighboring nodes: a successful fusion at $v$ which acts on the relevant $n$ qubits will result in a single, distributed, $n$-party entangled state shared by the $n$ neighbors.
Each node can perform a fusion on any number of successfully
entangled qubits, and each fusion succeeds with probability
$q_n$. This operation encompasses entanglement swaps, which are fusions on two
qubits. Nodes have knowledge of which qubits are involved in
successful links due to entanglement
heralding~\cite{link-layer-protocol, heralding}.

Analogously to a chain of entanglement swaps, fusions can be applied
across a ``network'' of repeaters. While a sequence of swaps extends
entanglement linearly, the effect of many fusions over a general
network is that entanglement ``spreads'' through the network; as long
as \textit{any} path of successful links and fusions exists between the
source and the destination, entanglement is distributed.

Figure~\ref{fig:multi-path-routing} illustrates a scenario in which fusion-based routing out-performs swapping-based routing.  Figure~\ref{fig:multi-path-routing}(a) shows an
example of single-path routing: a single path of channels is selected
to distribute entanglement, but one of the channels fails
to produce a link (Figure~\ref{fig:multi-path-routing}(b)), and entanglement distribution
is unsuccessful.  Figure~\ref{fig:multi-path-routing}(c) demonstrates
the utility of multi-path routing, which allows ``recovery
edges'' to be pre-selected to mitigate the effect of link failures.
Figure~\ref{fig:multi-path-routing}(d) shows that such a scheme is able to withstand the failure of the same link by using
a parallel recovery channel.  In this example, the local swap at node
$D$ fails, and distribution is ultimately unsuccessful.
Figure~\ref{fig:multi-path-routing}(e) demonstrates fusion, whereby
\textit{all} chosen channels comprise the routing ``path.''  The example shows
how fusion-based protocols can withstand failures of both link
establishment and local swaps, as 
either a successful path along ACDFB or along ACEFB would distribute entanglement; in this case the latter succeeded (Figure~\ref{fig:multi-path-routing}(f)).

Fusion-based protocols do not have to select the exact sequence of
nodes that will lead to end-to-end entanglement distribution
a-priori~\cite{pant-routing}.  Assuming sufficient connectivity,
fusion allows for attempting entanglement across ``all'' possible
paths simultaneously:
this observation forms the basis of a compelling result
in~\cite{Towsley}, which shows that in 2-D square grid networks, a
near unit entanglement distribution success rate is feasible,
regardless of distance, as long as $p$ and $q$ are beyond a critical
threshold.  Follow up work~\cite{multiplexed-Towsley} extends this
protocol for links with lifetimes that exceed a single time slot and
proposes a simple network partitioning scheme to improve the number of
distributed entanglements between a single S-D pair.  Kaur and
Guha~\cite{n-fusion-noise} study the square grid fusion protocol under
noisy link generation.
Sutcliffe and Beghelli \cite{multi-partite-routing} use fusions to
construct a protocol for multi-partite entanglement distribution.
\cite{qnet-percolation-analysis-1,qnet-percolation-analysis-2} analyze
the impact of network topology and connectivity on entanglement rates
over different distances for networks utilizing fusions.  Zeng
\etal{}~\cite{alg-n-fusion} incorporate fusions into a concurrent,
topology-agnostic routing protocol by generalizing the methods
in~\cite{Q-CAST}.

\newcommand{\chSet}{\ensuremath{\mathcal{C}}\xspace}
\newcommand{\eSet}{\ensuremath{\mathcal{E}}\xspace}
\section{Design\label{design}}

\begin{figure}
    \centering
\includegraphics[width=\columnwidth]{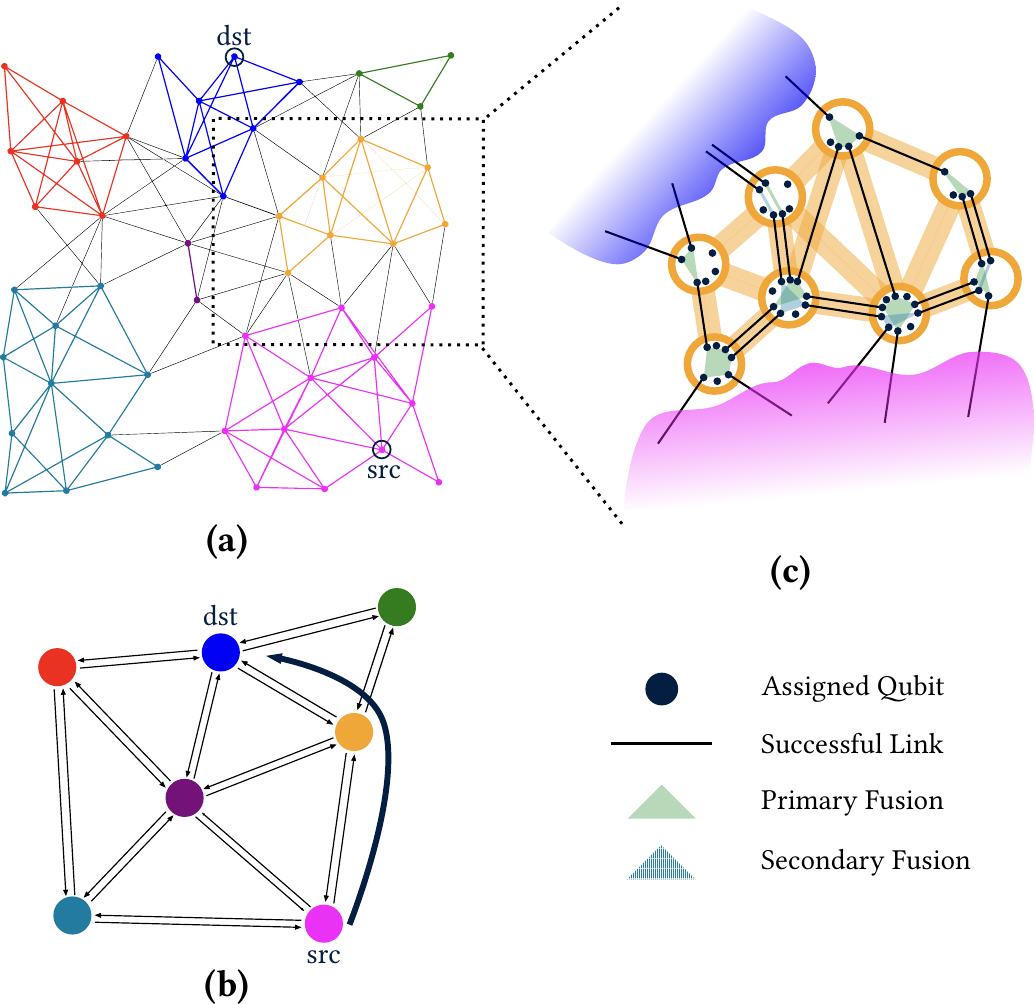}
\caption{Routing in \quarc.
  (a) The network is partitioned into
  \blobs. Nodes are colored by the \blob they belong to. 
  (b) Shortest path routing over \blobs. 
  (c) \Blob nodes assign qubits to
  channels, attempt link generation, and perform fusions to route
  entanglement from the source to the destination.
}
\label{routing}
\end{figure}

Figure~\ref{routing} sketches entanglement routing in \quarc at a high
level.
At any given time, the network is partitioned into a set of \blobs  (Figure~\ref{routing}(a)).  The input to \bRt is a set of source-destination
pairs (S-D pairs) of nodes that request an end-to-end entanglement.
Using Dijkstra's shortest path algorithm, \bRt selects an end-to-end
path over the graph induced by the \blobs (Figure~\ref{routing}(b));
here, the individual nodes and edges within the chosen \blobs define all possible
entanglement paths between the source and destination.  \bRt then assigns qubits to channels in a manner that distributes entanglement generation attempts  across all edges within the chosen path of \blobs in an approximately uniform manner. 
Each node then performs one or more fusions on its incident successfully entangled links to try to establish entanglement between the source and the destination.

Time in \quarc is divided into epochs.
Data about each cluster's effectiveness at distributing
entanglement is collected throughout each epoch.  Larger \blobs, with
more nodes, generally have a higher probability of being involved in a
successful entanglement, as these provide more potential entanglement
paths in our fusion-based scheme. The opposite is true for smaller \blobs.  While larger \blobs
increase the success probability of a given entanglement request, they
may also decrease \textit{overall} throughput since a \blob can only be
involved in the service of a single S-D pair at any given time.  (In the
degenerate case, the entire network may be a single \blob, at which
point, \bRt reduces to a protocol similar to that proposed
in~\cite{Towsley}).  Thus the basic idea of \bRt is to balance this
tradeoff: create \blobs just large enough to ensure end-to-end
entanglements can be established, while small enough such that multiple
entanglements can proceed simultaneously within the network.

\Blobs are evaluated and potentially reconfigured at the end of each
epoch.  In particular, if a \blob's rate of entanglement distribution
is deemed ``too low,'' its size is increased by merging it with some
number of nearby \blobs.  Conversely, if a \blob's rate is deemed
``too high,'' its size is decreased by splitting it into multiple
\blobs.

In the rest of this section, we describe details of how \blobs measure
their performance and how \blobs are re-configured, followed by a full
description of \bRt.

\paragraph{\Blob Entanglement Passing Rate}
Each time \quarc attempts to service an S-D pair, each \blob involved
in this entanglement distribution attempt records whether a successful path
of links and fusions was established from the previous
\blob (or source, if contained within the \blob) to the next \blob (or
destination).
The proportion of successful attempts is logged per epoch as the
\blob's \textit{entanglement passing rate}.

This process returns an estimate of the true expected entanglement
passing rate for each \blob.  This rate depends upon many factors,
including the size of a \blob, the \blob's internal connectivity, the
\blob's connectivity to adjacent \blobs, the S-D request distribution,
and link establishment rates under prevailing physical conditions.
The latter is particularly important, as there is an interaction
between how quickly physical conditions change 
and epoch length.  Epochs must be long
enough for the estimator to gather useful data, but also not so long
as to ignore material changes in physical parameters. It is possible that the \blob
structure be re-configured manually if physical parameters or the S-D request distribution
change drastically, but in \bRt, we rely entirely on the entanglement
passing rate measurements to reconfigure \blobs in response to these changes.

\subsection{\Blob Reconfiguration}

\begin{figure} 
  \centering
\includegraphics[width=.98\columnwidth]{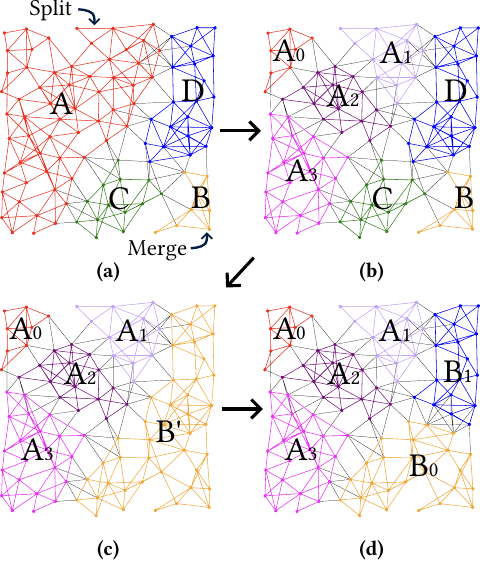}
\caption{\quarc \Blob reconfiguration: \blob A (red) opts to split,
  and \blob B (orange) opts to merge.  Panel (d) shows the final configuration.}
\label{sm}

\end{figure}

When should a \blob be split or merged? Consider a relatively large
\blob, e.g., A in Figure~\ref{sm}(a), that has high entanglement
passing rate.  Intuitively, it seems that this \blob could be split,
and the overall throughput in the network would increase as the split
\blobs would enable more concurrent routing requests and (ideally)
retain a high enough entanglement passing rate.  Splitting such a
\blob may be beneficial in a relatively small network, but maintaining
this \blob size may be required for high throughput in a larger
network, as S-D request path lengths grow (and end-to-end entanglement
establishment rate reduces exponentially).
The utility of a \blob depends not only on its size, but also
connectivity, both within and adjacent: the entanglement passing rate
metric captures all of these parameters in terms of how ``well'' a
\blob is performing locally.  \quarc uses a dynamic split threshold
that adds information about the network (its size, and optionally
topology), along with a \blob's current size and entanglement passing
rate, and only splits \blobs that exceed this threshold.

A similar reasoning motivates our merge scheme: merging small, low
entanglement passing rate \blobs will likely improve throughput, but
merging large \blobs, which usually have higher entanglement passing
rate, can reduce overall throughput.  Analogously, we choose a
dynamic merge threshold that again takes into account the same
factors.

Our basic approach follows the intuition that large \blobs (i.e.,
those containing many nodes) tend to have higher entanglement passing
rates because they provide more potential entanglement paths, and
vice-versa.  To split a \blob, we partition it into $k$ new \blobs using
the Girvan-Newman algorithm \cite{girvan-newman}, which
deterministically partitions a graph into a constant number of
densely-connected communities.  The Girvan-Newman algorithm is
particularly appropriate because it maintains high
connectivity for split \blobs, which in turn, results in higher
expected throughput for entanglement formation.  Our results show that
\quarc is relatively agnostic to small values of $k$ ($\in [2, 5]$);
we use $k=4$ by default. 

A possible merge procedure would be to merge $k$ adjacent \blobs.  We
opt for a somewhat more conservative approach, which enables a
finer-set of \blob arrangements, without drastically reconfiguring the
network every epoch.  In particular, to grow a \blob, we select two of
its neighbors and partition all nodes in these three \blobs into
two \blobs using the Girvan-Newman algorithm (see
Figure~\ref*{sm}~(c-d)).  This has the effect of
eliminating one \blob from the network, and the use of the
Girvan-Newman algorithm maintains well-connected \blobs.

Figure~\ref{sm} illustrates the process of \blob split and merge in
\quarc.  In this example, the large red \blob (labeled A) has opted to
split, and the orange \blob (labeled B)  has
opted to merge.  Panel (b) shows the result of a split (with $k=4$)
as output by the Girvan-Newman algorithm, splitting A into A0, A1, A2,
and A3.  To merge, \blob B identifies two neighbors, C and
D, and these three \blobs merge to form an
intermediate \blob B' in panel (c).  B' is then split into two, B0
and B1, in panel (d), completing the merge procedure.

\subsubsection{Threshold Selection}

Technically, a set of \blobs should be merged iff the \textit{incremental} percolation benefit from merging contributes more to
overall throughput than is lost due to the reduced number of
concurrent S-D requests that can be attempted.  Similarly, a \blob
should be split iff the reduction in the probability of establishing
entanglement for a given S-D pair due to reduction in percolation is
outweighed by the the gain in overall throughput due to increased
number of concurrent S-D pairs.

Percolation benefits have been studied extensively in regular
networks, including grid networks~\cite{LI20211}.  Percolation theory
identifies critical connectivity thresholds beyond which grid networks
provide S-D connectivity with high probability (and not otherwise).
This observation was key in the development of fusion-based routing
schemes~\cite{Towsley, multiplexed-Towsley} that were, in part, an
inspiration for \quarc.

\begin{figure}[t]

    \begin{subfigure}[t]{.48\columnwidth}
    \includegraphics[width=\textwidth]{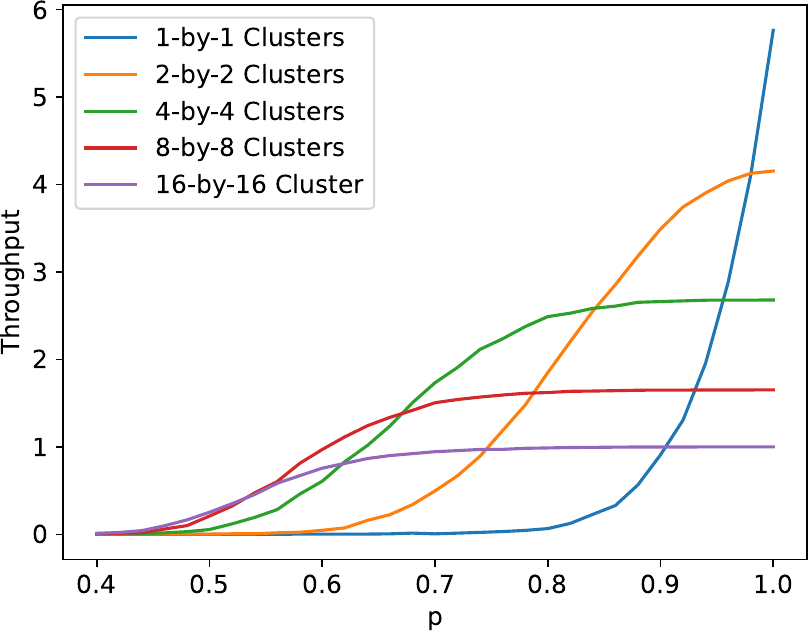}
    \caption{Comparison of static \blobbing with varying $p$}
  \end{subfigure}
  \hspace*{.01\textwidth}
  \begin{subfigure}[t]{.48\columnwidth}
    \includegraphics[width=\textwidth]{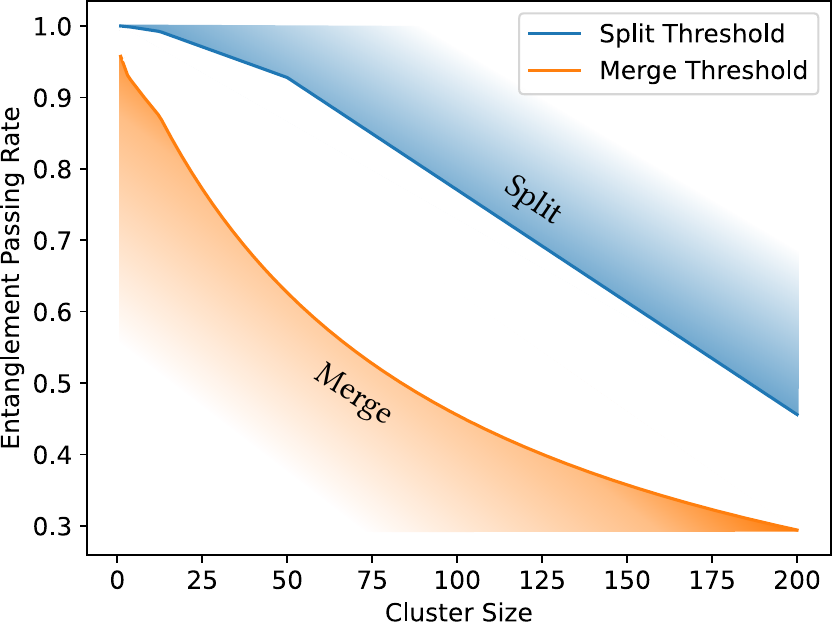}
    \caption{Splitting and merging thresholds for $n=200$}
  \end{subfigure}
  \caption{Threshold calculation.}
  \label{thresholds}
\end{figure}

Although existing theoretical percolation results apply only to a
specific set of networks (e.g.,~\cite{qnet-percolation-analysis-1,
  qnet-percolation-analysis-2}) and do not directly apply to \quarc's
network partitioning setting, we observe through simulation that such
a configuration still demonstrates discrete thresholding behavior.
Consider a $16\times16$ grid network with unit edge capacities, 4
qubits per node, and fixed $q$.  A natural set of \blob configurations
for this network is the set of square \blobs with side lengths that
are powers of two.
Letting $p\in[0,1]$ be constant and uniform for every channel, we can
measure the expected throughput of each \blob configuration through
simulation.  Sweeping over $p$, we find that the optimal configuration
indeed changes based on the value of $p$ with distinct transition
points (Figure~\ref{thresholds}(a)).  Similar patterns emerge as we
vary grid size.

Mapping these transition points to their corresponding average
entanglement passing rates gives us optimal merging and splitting
thresholds for 2-D grids.  We generalize these thresholds to \blobs of
arbitrary sizes (non powers of 2) by plotting our known thresholds as a
function of \blob size and linearly interpolating between \blob sizes
(Figure~\ref{thresholds}(b)).  We use the same procedure to generate
thresholds for other regular ($N\times N$) 2-D grids.  We linearly interpolate
$N\times N$ grid thresholds to compute thresholds for other network sizes.

\newcommand{\critP}{\ensuremath{p^*}\xspace}

\paragraph{Topology-specific Thresholds}
The 2-D grid derived thresholds can be used to refine thresholds for
arbitrary (families of) topologies.  For an arbitrary topology (say
G), we derive a custom threshold as follows: we simulate \quarc on 
G starting with the 2-D grid thresholds, with varying \textit{average}
value of entanglement generation probability $p$.\footnote{Note that in
non-grid topologies, the $p$ value changes by edge length.}  We find
the critical value \critP where using singleton \blobs provides the
same throughput as using \quarc with thresholds derived from the 2-D
grids.  We next simulate \quarc over G using \critP as the average
entanglement passing rate, and document the entanglement passing rates
$e_i$ for the \blobs \quarc creates at steady state.  We compute $G_t$
as the 75\%-percentile value of $e_i$ and set the topology-specific
threshold to be min($G_t$, $2D_t$), where $2D_t$ is the threshold
computed using the 2D-grid.  We use the 75\%-percentile value to be
conservative in splitting.

We use the same procedure using \critP and the average entanglement
passing rate of the singleton \blobs to set the merge threshold.  In
our evaluation, we present results from both 2D-grid thresholds
(scaled by network size) and using topology-specific thresholds as
described here.

\paragraph{Escaping local minima} \Blobs in quarc can reach a state
whereby the thresholds are insufficient for them to adapt, even though
the \blob-structure around has changed significantly.  To escape such
``local'' steady-states, we split a \blob $C$ of size $c$ if the
majority of its neighbor \blobs are less than size $c/k$.  Recall
that $k$ is the split constant provided to the Girvan-Newman algorithm (a \blob splits into $k$ \blobs).  We choose the $c/k$
neighbor size threshold because when \blob $C$ splits, it will create
\blobs of average size $c/k$.

\subsection{\bRt Protocol}

\bRt is initiated with a single \blob consisting of all nodes by
default.  At the end of each epoch, we examine the entanglement passing rate of
each \blob. 
Here, we employ the \blob configuration protocol in Algorithm~\ref{alg:reconfigure}.
Based on the thresholds described above, we mark each \blob for splitting if its entanglement passing rate is above the splitting threshold (or it meets the criterion for escaping a local minimum, see above) and otherwise mark it for merging if its entanglement passing rate is below the merging threshold (lines \ref*{line:to-split}-\ref*{line:to-merge}).
We split those \blobs marked for splitting as described above (lines \ref*{line:begin-split}-\ref*{line:end-split}).
We then consider the \blobs marked for merging in increasing
order of entanglement passing rate.
For each, we
choose the two neighbors 
which induce the smallest Kemeny
constant\footnote{We choose the Kemeny constant as a
measure of the connectedness of a component.  Other choices, such as
diameter, are also viable.}~\cite{Kemeny} (line \ref*{line:Kemeny}); if all relevant \blobs are unmodified, we merge the chosen
\blobs using the merge protocol described above and mark the new \blobs as modified (lines \ref*{line:if-unmerged}-\ref*{line:end-if-unmerged}).
If the \blob has exactly one neighbor (and it is unmodified), then we
directly merge the two \blobs into one.

\subsubsection{Routing Over \Blobs}

Within each time slot, we select routing paths over \blobs, assign
qubits to channels, attempt entanglement generation, and perform
fusions.  Figure~\ref{routing} shows the overall process, and we
refer to this figure in our description below.

\paragraph{Path Selection}

The S-D pairs in the request queue (e.g., in
Figure~\ref{routing}(a), nodes \textit{src} and \textit{dst} request entanglement) are prioritized by their arrival times
in order to minimize starvation.  For each pair, a path of \blobs is
selected using Dijkstra's algorithm over the graph induced by the
\blobs, where the weight of each (directed) edge is defined to be the
size of the \blob on the tail end of the edge divided by the number of
quantum channels between the two \blobs (Figure~\ref{routing}(b)). This
edge weighting prioritizes routes with high connectivity between
\blobs, while de-emphasizing routes that use excessive network
resources.  The selected \blobs are removed from the graph and the
process is repeated for the next S-D pair. If the remaining resources
do not allow a path between an S-D pair, that pair is skipped (in the
current time slot).  Any skipped pair ages, and is likely to be
selected earlier in subsequent rounds.

\paragraph{Qubit Assignment}

Suppose the \blob path $C_0, C_1, \dots, C_k$ is selected for a given S-D
entanglement request.  In \quarc, we do not assume that nodes can
necessarily allocate qubits to all possible channels (i.e., nodes may
be memory limited), and thus, \quarc needs to find a feasible
assignment of qubits to channels.  Qubit assignment proceeds in two
steps: first, \quarc creates a union of all edges (\eSet) and a
union of channels (\chSet) that are within $C_i$ or between $C_i$ and
$C_j$, $i\neq j \in [0, k]$.  
Each edge $e_\ell \in \eSet$ is assigned a priority $a_\ell \in [0,1]$ picked uniformly at random. Channels within each edge are assigned a priority $(a_\ell + i)$, where
$i$ is an integer index for the
channel, starting from 0 for the first channel, 1 for the second, and
so on.  Channels are then sorted by priority and qubits are assigned
sequentially in this sorted order, as long as each end of a channel
has a free qubit remaining.

This method of allocating qubits ensures that, if possible,
at least one qubit is allocated to every edge
before any edge is assigned a second, and so on.  However, this allocation of qubits is also
essentially ``random,'' in that it does not take into account factors
such as the shortest path between S and D, channel width of edges,
number of parallel S-D paths that have received qubits, the values of
$p$ per edge or $q$ per node, etc.  This assignment serves to
disperse qubits across the \blob.
We discuss more enhanced qubit assignment strategies that can be used
to optimize different potential metrics in Section~\ref{discuss}.

\addtocounter{footnote}{1} 
\begin{algorithm}[t]
\small
    \caption{Adaptive \Blob Reconfiguration} 
    \begin{flushleft}
        \Input $G = \langle V,E,C \rangle$, current \blobs $C=\{c_1,\dots,c_{|C|}\}$, sorted by entanglement passing rates $\{r_{c_1},\dots,r_{c_{|C|}}\}$, merging and splitting thresholds $m$ and $s$, respectively, constant $k$.\\
        \Output New set of \blobs (partitioning of $V$)
    \end{flushleft}
    \begin{algorithmic}[1]
        \State $S \gets \{c\in C \mid r_{c}\geq s(|c|)\}$ \Comment{\Blobs to split} \label{line:to-split}
        \State $S \gets S\cup \{c\in C \mid \text{majority of \textit{c}'s neighbors have size}<|c|/k\}$
        \State $M \gets \{c\in C\setminus S \mid r_{c}\leq m(|c|)\}$ \Comment{\Blobs to merge} \label{line:to-merge}
        
        \For {$c\in S$} \label{line:begin-split} 
            \State $d_1,\dots,d_k \gets \texttt{Girvan-Newman}(c,k)$\\
            \Comment{Split $c$ into $k$ components}
            \State $C\gets (C\setminus \{c\})\cup \{d_1,\dots,d_k\}$ \label{line:end-split} 
        \EndFor

        \State $K\gets\emptyset$ \Comment{Set of merged \blobs}
        
        \For {$c\in M$ such that $c$ has neighbors}

            \State $x_1, x_2 \gets c$'s two neighbor \blobs\footnote[\thefootnote]{This footnote doesn't appear, make sure the number lines up with the one in the text.} that minimize induced Kemeny constant \label{line:Kemeny}
            \If{$c,x_1,x_2\notin K$} \label{line:if-unmerged}
                \State $d_1, d_2 \gets \texttt{Girvan-Newman}(c\cup x_1 \cup x_2, 2)$\\
                \Comment{Split three \blobs into $2$}
                \State $C\gets (C\setminus \{c,x_1,x_2\})\cup \{d_1, d_2\}$
                \State $K\gets K \cup \{c, x_1,x_2,d_1,d_2\}$ \label{line:end-if-unmerged}
            \EndIf
        \EndFor    
        \\
        \Return{$C$}
    \end{algorithmic} 
\label{alg:reconfigure}
\end{algorithm}

\footnotetext{In the
  edge case where $c$ has exactly one neighbor $x$, we directly merge
  $c$ and $x$ into one \blob in a manner analogous to lines
  \ref*{line:if-unmerged}-\ref*{line:end-if-unmerged}.}

\paragraph{Fusion Protocol}

Finally, all nodes follow a local fusion protocol similar to the
protocol presented in~\cite{multiplexed-Towsley}.
Each node attempts entanglement generation according to its qubit
assignments. All channels are given predetermined (arbitrary) ids. The
node selects the successful link with the lowest id from each incident
edge and attempts a fusion operation on these selected channels  (labeled primary fusions in Figure~\ref{routing}(c)).  It
then repeats this process on the remaining links until no links
remain (secondary fusions in Figure~\ref{routing}(c)).  If a selection of links would leave only one link remaining,
that last link is included in the previous fusion attempt.  We note
that the primary fusions alone give the same result as fusing all
links in the selected \blobs at once, so the inclusion of secondary fusions only serves to
increase the entanglement rate.  All nodes send their fusion
measurement results to the destination node, which computes whether or
not the entanglement distribution was successful and applies the usual
qubit corrections~\cite{multiplexed-Towsley, Towsley}.

\newcommand{\qcast}{Q-CAST\xspace}
\newcommand{\anf}{ALG-N-FUSION\xspace}
\newcommand{\ANF}{\anf}
\section{Evaluation\label{eval}}

We present an evaluation of \quarc, including its ability to adapt,
and a comparison to current state-of-the-art quantum routing
protocols.

\subsection{Methodology\label{methodology}}

\quarc's simulation code was custom written in Python, and allows the
following parameters to be modified:
\begin{itemize}
\item Network topology: number and location of nodes, location and widths of edges
\item Physical parameters: number of qubits per node, entanglement generation
probability ($p$) per channel, and fusion success
  probability ($q$)
\end{itemize}

Once the physical parameters
are set, the simulator can be configured to generate entanglement
requests from configurable distributions. We generate S-D pairs
uniformly at random unless otherwise noted.  Once an S-D pair requests
entanglement, the request remains in a global FIFO request queue until
it is satisfied.  The size of the request queue is configurable, and
is set to 10 in our experiments unless otherwise noted.  New
entanglement requests are generated and appended to the request queue
as existing ones are satisfied, which means that the request queue
always contains ten requests (by default).  Our choice of retaining
failed entanglement requests differs from that of prior
research~\cite{Q-CAST, alg-n-fusion}, where failed requests are
discarded and not tried again.  It is not clear whether it is
reasonable to simply discard unsatisfied entanglement requests; this
may depend on the specific quantum application.  For example, if the
task is to generate genuine quantum entanglement across the
network~\cite{genuine-entanglement-pan-2023}, requests to entangle
distant nodes cannot be simply discarded.  We note, however, that it
is also not clear that a satisfying a request after a long amount of
time is necessarily useful; this again may be application-dependent.
While \quarc prioritizes requests that were not satisfied in previous
time slots, we do not impose this restriction on other protocols,
which are allowed to optimize outstanding requests without 
limitation from our simulation framework.  We ensure that the size of
the request queue is never a limiting factor in any of the results
presented below.

\begin{figure*}[t]
\centering

\begin{subfigure}[t]{.95\textwidth}
    \raisebox{.5\height}{\includegraphics[width=\textwidth]{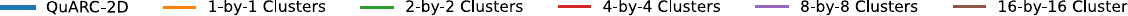}}
\end{subfigure}
\bigskip
\begin{subfigure}[t]{.25\textwidth}
    \includegraphics[width=\textwidth]{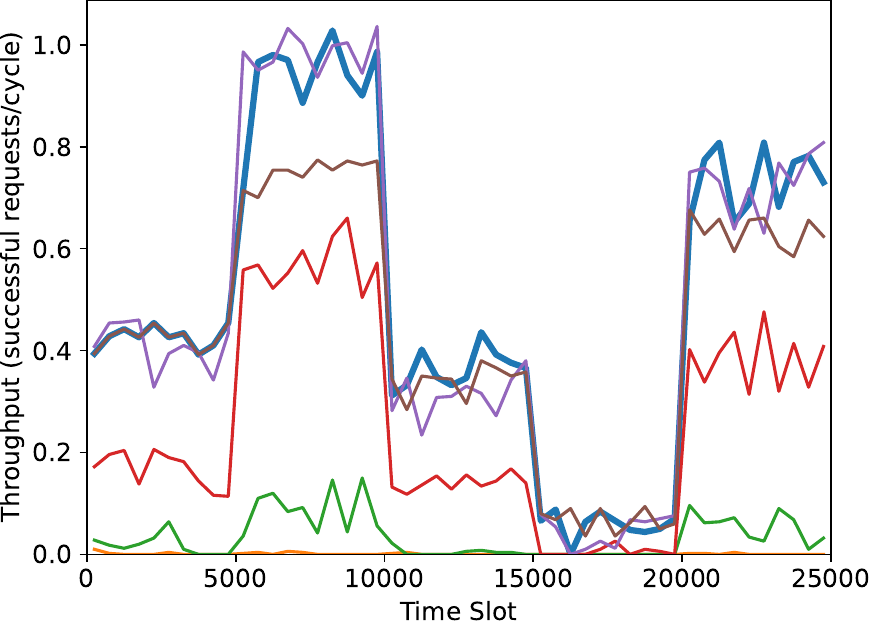}
    \caption{p and q both change:\\($p$,$q$)~$\in[(.7,.8),(.6,1.0),(.8,.7),\\\hspace*{0.4in} (.5,.9),(.9,.7)]$}
\end{subfigure}
\hspace*{.01\textwidth}
\begin{subfigure}[t]{.25\textwidth}
    \includegraphics[width=\textwidth]{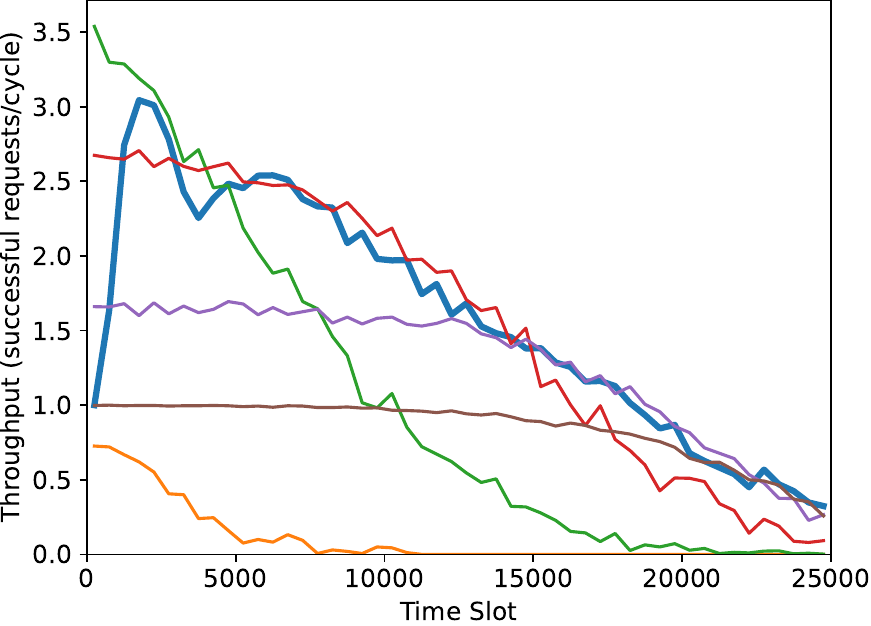}
    \caption{p decreases from 0.9 to 0.5 by 0.01 every 400 time slot}
\end{subfigure}
\hspace*{.01\textwidth}
\begin{subfigure}[t]{.25\textwidth}
    \includegraphics[width=\textwidth]{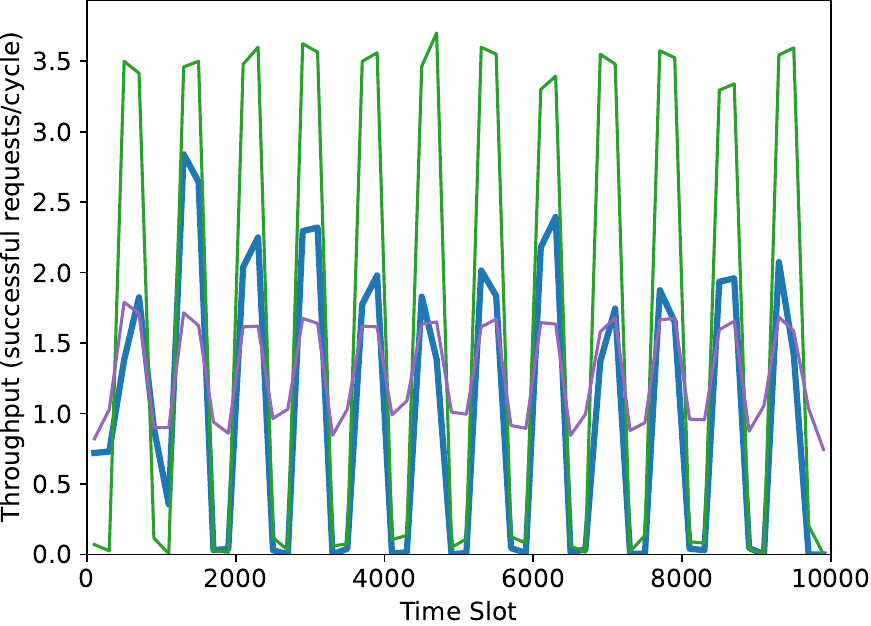}
    \caption{p oscillates between 0.6 and 0.9 every 400 time slot}
\end{subfigure}
\hspace*{.01\textwidth}
\begin{subfigure}[t]{.184\textwidth}
    \includegraphics[width=\textwidth]{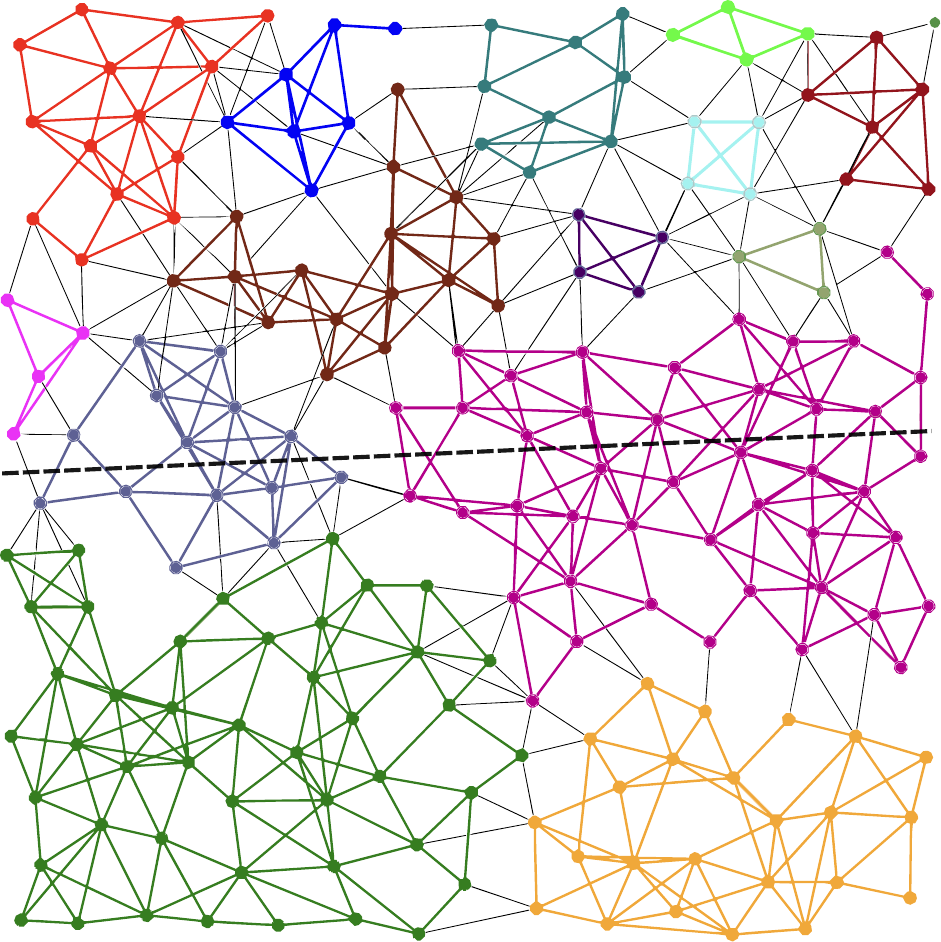}
    \caption{Spatial Adaptation: $p = 0.6$ in the upper half, $p=0.3$ in the lower half}
\end{subfigure}

\caption{\quarc adaptation with changing physical parameters ($p$ and
  $q$); (a-c) $16\times16$ grid network. (a) Sudden parameter shifts. (b) Gradual parameter decay. (c) Sharp parameter oscillation. (d) Spatially-varying parameters.}
\label{fig:learning}
\end{figure*}

Our evaluation of \quarc is designed to address two broad questions:
(1) How well can \quarc adapt to changing physical parameters and (2)
How does \quarc's performance compare to state of the art entanglement
routing protocols?  For the first, we present results comparing
\quarc's dynamic \blob configuration to fixed \blob arrangements,
while varying $p$ and $q$ over time and space.  For the latter, we
compare \quarc to \qcast~\cite{Q-CAST} (a concurrent swapping-based
protocol) and \anf~\cite{alg-n-fusion} (a concurrent fusion-based
protocol).  Our primary measures of performance are throughput (number
of successful entanglements over time), latency (time before
entanglement is successful) and starvation rates (fraction of
entanglement requests that are unsuccessful).  We note that there is a
lack of consensus in how throughput ought to be measured in quantum
settings: for instance, in many quantum routing
evaluations~\cite{Q-CAST,alg-n-fusion,REPS,fragmentation-Q-CAST}, 
if a S-D request is satisfied
with three parallel entanglements, then the throughput for that
particular request is considered to be 3.
Such a configuration is useful if the application required three
entanglements.  Papers that focus on quantum network tasks, e.g.,
quantum distributed sensing network~\cite{Zhang_2021} or
entanglement-based long-baseline
interferometry~\cite{PhysRevLett.109.070503}, on the other hand, state
that establishing all requested S-D entanglements is more important
than counting the multiplicity of entanglements for each S-D pair.
In our results, we count a request as satisfied if at least one S-D
entanglement is achieved, and, by default, do \textit{not} count the
number of parallel entanglements in the measure of throughput.  (We do
discuss results with the ``aggregate throughput'' measure in select
experiments.)  A more realistic model would augment each request with
the desired number of entanglements, an extension which would likely
require changes to the design of \quarc and other protocols, which we
leave as future work.

\subsection{\quarc Adaptation}

The initial question we wish to investigate is whether \quarc can
efficiently adapt the underlying \blob structure as network parameters
evolve.  
For this experiment, we use a 2-D grid network, as 2-D square \blobs form a uniform configuration with high inter- and intra-\blob connectivity, and therefore constitute a natural baseline.

Figure~\ref{fig:learning}(a-c) shows \quarc's performance on a 16x16
grid, with a single channel per edge and no qubit limitations, as
physical parameters $p$ and $q$ change.  The $x$-axis is simulation
time, and \quarc is started at time 0 with the default configuration
of a single \blob. The \blob structure is not manually changed as $p$
or $q$ change.  The other curves on the figure correspond to fixed
\blob configurations, which provide high throughput for some
combinations of $p$ and $q$, but not others.  Note that the 16x16
\blob configuration corresponds to a single \blob; on this topology,
this configuration reduces to the 4-GHZ protocol described
in~\cite{Towsley}.

Figure~\ref{fig:learning}(a) corresponds to a situation where both $p$
and $q$ change abruptly but over long time intervals (every 5000 time
slots) compared to \blob reconfiguration time (500 time slots).
\quarc is able to reconfigure the \blobs efficiently to track physical
parameters, and reaches a stable state with performance equivalent to
the best static \blob configuration (for that $p/q$) relatively
quickly.  We also experimented with changing $p$ and $q$ separately,
and the results (not plotted) showed that \quarc is able to adapt
efficiently.

Figure~\ref{fig:learning}(b)
shows a case where $p$ decreases from 0.9 to 0.5 by 0.01 every 400
time slots.
This situation is designed to reflect a case where link
performance degrades over time, for instance due to changes in
temperature or loss of synchronization between nodes~\cite{Fang_2023,boston-qnet}.
Note that in this scenario, network parameters change more
frequently (every 400 time slots) than \quarc is able to reconfigure (500 time slots).
\quarc is again able to provide high
throughput, essentially equivalent to the best fixed \blob
configuration.  \quarc's performance tends to lag slightly behind the best configuration, but is able to eventually find a suitable arrangement.
Other experiments, not plotted, show that \quarc is able to similarly track
increasing $p$.

Figure~\ref{fig:learning}(c) explores an extreme scenario in which the
network is unstable.  Here, $p$ oscillates between 0.6 and 0.9 every
400 time slots, and \quarc reconfigures every 500.  As expected,
\quarc is unable to track such rapid changes, but does provide some
throughput when the entanglement probability rises.  In practice, the
epoch time used to initiate reconfiguration would have to be carefully
selected based on how physical parameters evolve.

In addition to temporal adaptation, Figure~\ref{fig:learning}(d)
illustrates how \quarc's \blobbing also adapts to spatial disparities.
The figure shows the resulting \blob configuration in a scenario on a
200 node Waxman network~\cite{Waxman} where channels in the top half
of the topology have $p=0.6$, and channels in the bottom half have
$p=0.3$ (other parameters follow the reference setting, described
below). \quarc's adaptation is able to detect this spatial disparity
and, after convergence, form larger \blobs in the bottom half of the
network compared to the top.

\subsection{Comparison of \quarc vs. existing protocols}

\begin{figure*}[t]
\centering

\begin{subfigure}[t]{\textwidth}
    \centering
    \raisebox{.4\height}
    {\includegraphics[width=.5\textwidth]{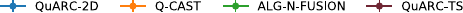}}
\end{subfigure}
\bigskip
\begin{subfigure}[t]{.232\textwidth}
    \includegraphics[width=\textwidth]{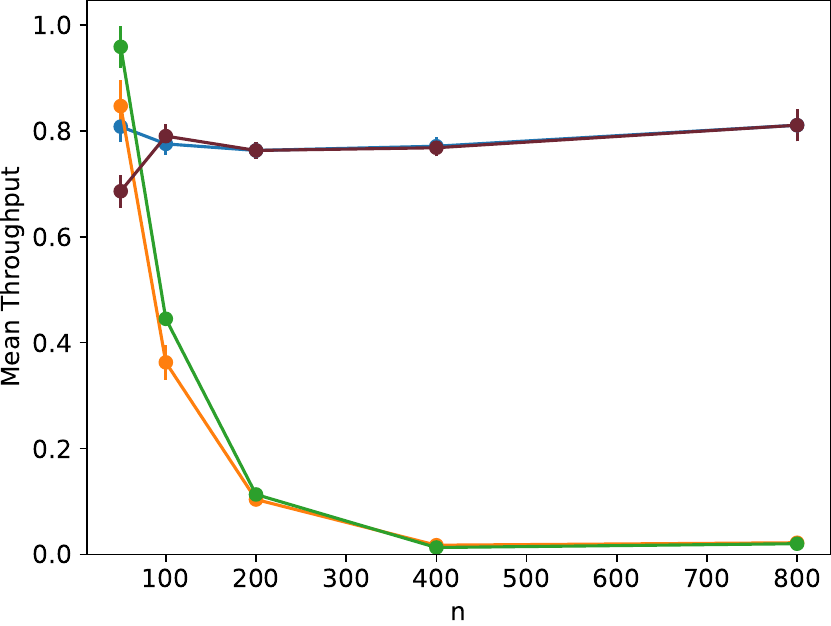}
    \caption{Throughput vs. Network Size ($E_p=0.3$)}
\end{subfigure}
\hspace*{.01\textwidth}
\begin{subfigure}[t]{.232\textwidth}
    \includegraphics[width=\textwidth]{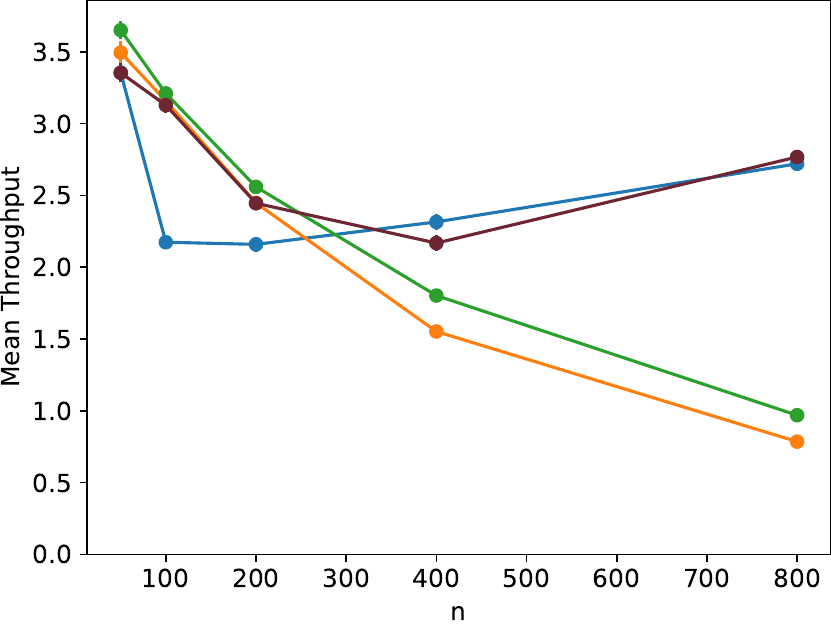}
    \caption{Throughput vs. Network Size ($E_p=0.6$)}
\end{subfigure}
\hspace*{.01\textwidth}
\begin{subfigure}[t]{.232\textwidth}
    \includegraphics[width=\textwidth]{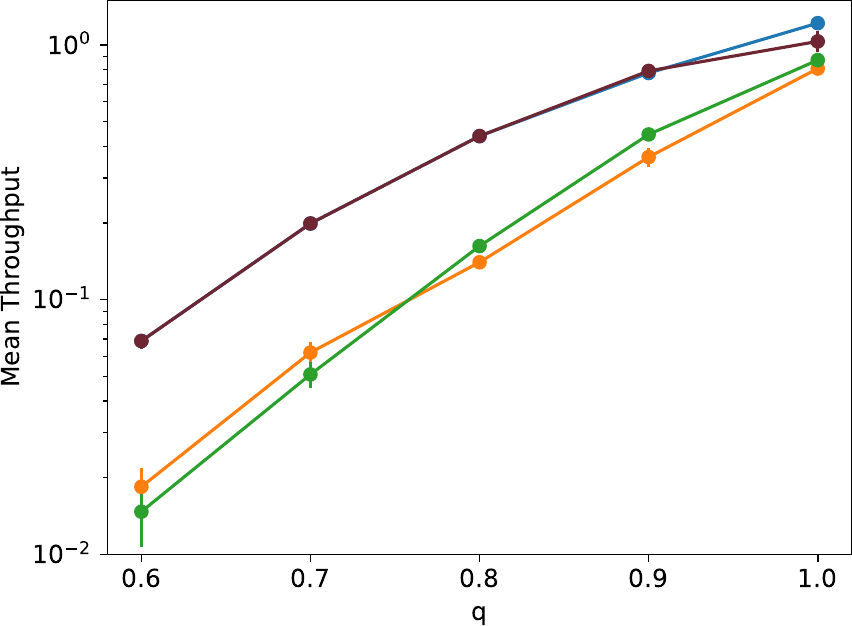}
    \caption{Throughput vs. Fusion Success Probability ($E_p=0.3$)}
    \label{fig:tp-vs-q}
\end{subfigure}
\hspace*{.01\textwidth}
\begin{subfigure}[t]{.232\textwidth}
    \includegraphics[width=\textwidth]{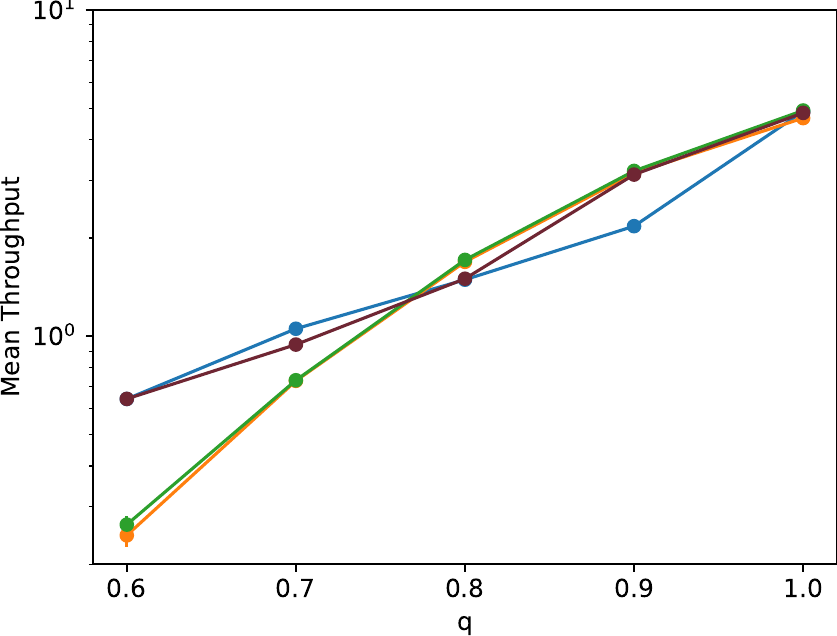}
    \caption{Throughput vs. Fusion Success Probability ($E_p=0.6$)}
\end{subfigure}

\caption{Throughput comparison with varying physical parameters and network size.}
\label{throughput}
\end{figure*}

In this section, we compare \quarc to two state-of-the art quantum
routing protocols, \qcast~\cite{Q-CAST} and \ANF \cite{alg-n-fusion}.
Unlike \quarc, both \qcast and \anf utilize a-priori knowledge of all
physical parameters; in particular, the values of $p$ per channel and
$q$ per node. For these results, we follow the same topology
generation method using the Waxman model~\cite{Waxman} described in
\cite{Q-CAST}, and also used in \cite{alg-n-fusion}. As in~\cite{Q-CAST}, we give each topology
an average degree of 6 and we assign each node a number of qubits
picked uniformly at random from the range $[10,14]$ and each edge a
width picked uniformly at random from the range $[3,7]$. We follow the
convention of denoting the average value of $p$ across all channels as
$E_p$.  Each plotted point is an average of ten different simulation
runs, with error bars showing the standard deviation of the mean.

\paragraph{Protocol Implementations}
For these results, we used the simulator described in
Section~\ref{methodology} for \quarc.  For \qcast, we used the
publicly available simulation code~\cite{Q-CAST-source-code} written
in Kotlin.  We did not change the provided code except to
implement our request queueing (if necessary) and data collection.
There was no publicly available code for \ANF: we implemented \ANF in
Kotlin, based upon the description in the original
reference~\cite{alg-n-fusion}.  All code used in these evaluations
is publicly available~\cite{quarc-source-code}.

\paragraph{Throughput}
The throughput of each protocol under varying conditions is shown in
Figure~\ref{throughput}.  Each plot shows throughput for a fixed value
of $E_p\in\{0.3, 0.6\}$, with other parameters set as in~\cite{Q-CAST}
unless otherwise noted.

Figure~\ref{throughput}(a) shows the throughput of \quarc, \qcast, and
\ANF as a function of the number of nodes $n$ in the topology when $E_p=0.3$.  
Both \quarc-2D and \quarc-TS demonstrate the utility of \blobbing 
in this more constrained setting. At lower values of $E_p$,
the throughput of \qcast and \ANF rapidly diminish to essentially zero
with increasing network size, as most requests are starved, whereas
both variants of \quarc are able to take advantage of the high path
diversity inherent in \blobbing to sustain S-D entanglement success
rates.  

Figure~\ref{throughput}(b) shows throughput as a function of network size in the more generous setting where $E_p=0.6$.
In the
\qcast reference setting ($n=100$, $E_p=0.6$), \qcast and \ANF
outperform \quarc-2D (\quarc with 2D-Grid thresholds and no knowledge
of qubit availability).  In this setting (and with $n=50$), average
path lengths are short, and the probability of a successful
entanglement on a given edge is high ($1-(1-0.6)^5 = 0.990$, assuming
average channel width of five).  This high entanglement generation
probability coupled with large number of available qubits leads to a
resource-rich environment, enabling \qcast and \ANF to generate a
comparatively high number of S-D entanglements.  In such an
environment, \quarc-2D is also able to generate a high entanglement
rate, but the underlying \blobbing throttles throughput as it reduces
the number of S-D pairs that \quarc-2D is able to attempt
simultaneously.  This disparity is eliminated when \quarc is run using
topology-specific thresholds (\quarc-TS): the new thresholds, that are
aware of both the underlying topology and qubit availability, increase
the probability of splitting, resulting in higher throughput compared
to \quarc-2D.

Despite their differences, a common picture emerges across both
Figures~\ref{throughput}(a-b) as network sizes increase: the
throughput of \qcast and \ANF decrease dramatically while the
throughput of \quarc is relatively stable.  As network sizes increase,
so do average path lengths, and the chance of a successful S-D
entanglement decreases exponentially, even though individual edges
still have a very high entanglement probability (about 0.99, as per
above), and the chance of successful entanglement swap at a node is
also high (0.9).  Both \quarc variants, on the other hand, are
relatively insulated from the increase in network size, as they can
reconfigure the \blob structure to account for increased path lengths.
As network sizes increase, both \quarc protocols automatically use
larger \blobs to effectively maintain a reasonable request success
rate.  (In results not plotted, we see that the \textit{number} of
\blobs created by \quarc protocols remain relatively stable as network
sizes increase, leading to increased \blob sizes for larger networks.)

\qcast and \anf are similarly affected if the value of $q$
changes.  Figures~\ref{throughput}(c-d) plot throughput ($y$-axis, log-scaled)
for all three protocols for $E_p\in\{0.3, 0.6\}$ as $q$ varies
($x$-axis) on the 100-node \qcast reference topology.  
As $q$ decreases, end-to-end entanglement probabilities decrease
exponentially for both \qcast and \anf, while both \quarc variants are
better able to compensate by creating larger \blobs.  While this
result plots throughput performance for a 100 node graph, \quarc's
relative advantage increases substantially for larger networks (not
plotted).

We discuss a few experiments not plotted due to space constraints.
As discussed in Section~\ref{methodology}, an alternate measure of
throughput accounts for the number of parallel entanglements achieved,
which can linearly increase throughput for successful entanglements.
We refer to this measure as \textit{aggregate throughput}.
Experiments show that trends in aggregate throughput are similar to
those in Figures~\ref{throughput}(a-b), but with a modest relative
boost for Q-CAST under favorable network conditions (small network
size and high entanglement probability).

\quarc's thresholds are derived using simulations that model a uniform
distribution of S-D pairs requesting entanglement.
We have explored \quarc's sensitivity to request distribution,
including modeling a ``worst'' case scenario for \quarc (least
homogeneous request mix) by using a bimodal request distribution in
which 50\% of requests have a hop-distance that is 25\% of the network
diameter, and 50\% of requests have a hop-distance that is 75\% of the
network diameter.
We find that \quarc variants are also relatively unaffected by changes
in request distribution, and maintain their advantage as network sizes
increase, or as entanglement probabilities decrease.

Overall, these results are instructive, as they show:
\begin{itemize}
    \item \quarc is able to maintain throughput performance as the
      network size increases across a wide range of  $p$ and $q$,
      as well as under non-uniform traffic patterns.
      
    \item The generic 2D-threshold version of \quarc can provide
      higher throughput than existing protocols on larger networks or
      with constrained values of $p$; topology-specific 
      \quarc maintains this advantage, and is able to better use
      network resources to perform on par with multi-path entanglement
      passing algorithms in smaller networks with high values of $p$
      and $q$.
      
    \item The performance of \qcast and \ANF is coupled to parameters
      such as entanglement generation probability $p$, fusion
      probability $q$, and network size.  Performance degrades rapidly
      with network size or reduction in $p$ or $q$ values; as network
      sizes increase, in order to maintain performance, these protocols 
      {\em require\/} that $p$, $q$ also increase, or that
      channel widths increase.
\end{itemize}

\paragraph{Latency and Starvation}

Along with throughput, latency (time before entanglement is
established) and starvation (number of S-D requests unfulfilled) are
important measures of performance.  In the results presented above,
requests linger until satisfied, but it is unclear whether quantum
applications can make use of entanglements much later than originally
requested.  Evaluations of previous
protocols~\cite{Q-CAST,alg-n-fusion} discard S-D requests if they
cannot be satisfied immediately (one time slot), which leads to these
protocols serving many low-hop count requests and essentially starving
all others.  Here, we explore the interaction between request latency,
starvation rate, and S-D distance.  For this set of results, we use a
200 node Waxman topology, $E_p=0.5$, $q=0.9$. We choose this setting
because all three protocols achieve similar throughput in this
setting.  The \quarc results here use the generic 2D-grid
thresholds, and are representative of using the topology-specific
thresholds.

\begin{figure}[t]
\begin{subfigure}[t]{\columnwidth}
    \raisebox{.5\height}{\includegraphics[width=\textwidth]{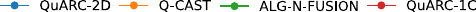}}
\end{subfigure}
  
    \begin{subfigure}[t]{.48\columnwidth}
    \includegraphics[width=\textwidth]{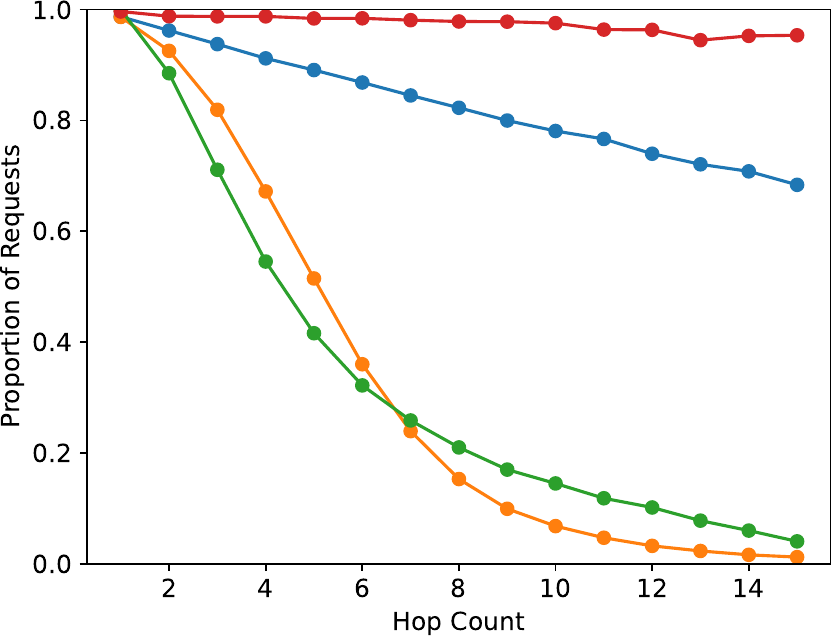}
    \caption{Starvation: proportion of successful entanglement distribution attempts by distance}
  \end{subfigure}
  \hspace*{.01\textwidth}
  \begin{subfigure}[t]{.48\columnwidth}
    \includegraphics[width=\textwidth]{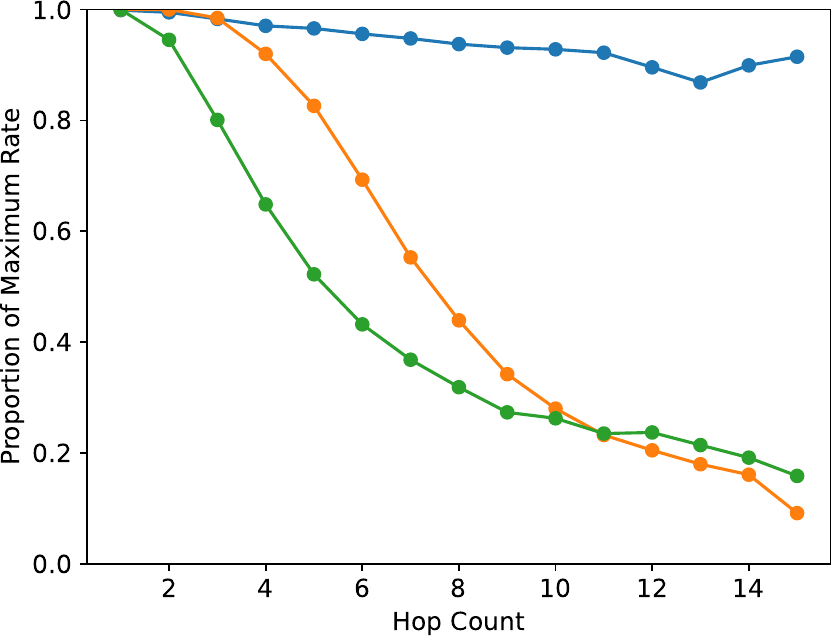}
    \caption{Resource allocation bias: realized proportion of maximum success rate}
  \end{subfigure}
  \caption{Protocol fairness vs request distance}
  \label{starve}
\end{figure}

Figure~\ref{starve}(a) depicts the bias in request fulfillment
as a function of distance.  The $y$-axis shows the proportion of
attempted requests that were successfully filled.  Both \qcast and
\anf show a much stronger bias compared to \quarc in their ability
to serve long requests, succeeding in fewer than 20\% of requests
longer than 8 hops.
\quarc also experiences
somewhat of a bias with distance; however this is by design. The red
curve labeled ``\quarc-1C'' represents the highest achievable success
rate when \quarc uses a single \blob. However, \quarc makes the
decision to use multiple \blobs in order to serve multiple concurrent
S-D pairs, at the expense of a modest amount of distance-independence.

Figure~\ref{starve}(b) illustrates the effects of resource allocation bias. For each protocol and hop count, the plotted value is the ratio of the realized success rate (as shown in Figure~\ref{starve}(a)) to the maximum success rate.
We measure the maximum success rate by supplying each protocol with a single S-D request at a time, isolating its fundamental ability to fill each request.
Both \qcast and \anf
preferentially attempt to serve shorter requests, resulting in a significant reduction in efficiency of filling long-distance requests compared to their maximum capability.
Meanwhile, \quarc's FIFO queue prioritization allows for little choice in request selection, resulting in minimal bias or loss of efficiency when servicing multiple concurrent requests.

\section{Discussion\label{discuss}}

We discuss aspects of \quarc we have not covered, including
ways in which \quarc's adaptation can be improved.

\subsection{\quarc Adaptation}

As evaluated, \quarc's adaptation scheme does not use any time-varying
global information.
We have experimented with providing \quarc with more global
information.  In particular, we have considered schemes which incorporate overall
(global) success rate in addition to entanglement passing rate.
Providing access to the global success rate did not result in substantial
performance gains, and we opted for the simpler \blob-local metric.

It is likely that providing access to physical parameters, such as
instantaneous $p$/$q$ values (as is assumed in \qcast and \anf),
requested S-D path length distribution, etc., can lead to more
effective \blob structures.  Such a design needs to factor in the
actual overhead of characterizing these
parameters~\cite{benchmark-quantum,NIST-qnet-metric} as well as the
control overhead of distributing such information across the
network.

Similarly, more advanced qubit assignment schemes may prove beneficial for throughput, but at the cost of additional overhead. \quarc's approach
allocates qubits in a distributed manner to try to
maximize percolation benefit.
However, it may be worthwhile to consider a scheme that allocates qubits on a request-specific basis to take advantage of specific topological features between a particular S-D pair.
A more systematic study
of this tradeoff is part of our future work.

\subsection{Implementing \quarc}

While our focus in this paper has been to introduce the idea of \blobbing in the
context of quantum routing, we have also designed and implemented a fully distributed version of \quarc that resembles hierarchical link-state
routing.
Here, each \blob elects a leader through an election protocol, and the leader collects
information about entanglement passing rate. The leader can then decide to initiate a split, or to initiate a merge by communicating the desire to merge with neighboring leaders.
Individual S-D requests are
source-routed (via the known \blob topology)
in coordination with \blob leaders.
The development of the necessary
sub-protocols for robust leader election, leader-leader communication,
time synchronization for initiating merges, and ordering of requests to be served
was informed by corresponding classical concepts, which have been well-studied.

\section{Conclusion\label{conc}}

We introduce \quarc, a new quantum routing protocol that uses dynamic
\blobbing to simultaneously provide high throughput while decoupling
performance from varying entanglement generation rates, swapping/fusion
probabilities, and network size.  \quarc does not use time-varying
global information for \blobbing decisions, and our simulation-based
evaluation shows that \quarc's performance, both in terms of
throughput, entanglement success rates, and request fulfillment
latency, is robust with respect to dynamic changes in physical
parameters.

\bibliographystyle{IEEEtran}
\bibliography{reference}

\end{document}